\documentclass[twocolumn,showpacs,preprintnumbers,amsmath,amssymb]{revtex4}

\usepackage{graphicx}
\usepackage{dcolumn}
\usepackage{bm}

\begin{document}


\title{Measurement of the electric fluctuation\\ spectrum of magnetohydrodynamic
turbulence}

\author{S. D. Bale}
\altaffiliation{Department of Physics and Space Sciences Laboratory, University of California, Berkeley}
\author{P. J. Kellogg} 
\altaffiliation{School of Physics and Astronomy, University of Minnesota, Minneapolis}
\author{F. S. Mozer}
\altaffiliation{Department of Physics and Space Sciences Laboratory, University of California, Berkeley}
\author{T. S. Horbury}
\altaffiliation{The Blackett Laboratory, Imperial College, London, UK}
\author{H. Reme}
\altaffiliation{CESR, Toulouse, France}
\date{\today}

\begin{abstract}
Magnetohydrodynamic (MHD) turbulence in the solar wind is observed to show the 
spectral behavior of classical Kolmogorov fluid turbulence over an inertial subrange
and departures from this at short wavelengths, where energy should be dissipated.
Here we present the first measurements of the electric field fluctuation spectrum over
the inertial and dissipative wavenumber ranges in a $\beta \gtrsim 1$ plasma.  
The $k^{-5/3}$ inertial subrange is
observed and agrees strikingly with the magnetic fluctuation spectrum; the wave phase speed
in this regime is shown to be consistent with the Alfv\'en speed.  At smaller wavelengths
$k \rho_i \geq 1$ the electric spectrum is softer and is consistent with the expected
dispersion relation of short-wavelength kinetic Alfv\'en waves.  Kinetic Alfv\'en waves
damp on the solar wind ions and electrons and may act to isotropize them.  This effect may explain the
fluid-like nature of the solar wind.
\end{abstract}

\maketitle

Turbulence is ubiquitous in astrophysical plasmas; turbulent processes are thought
to play a role in cosmic ray and energetic particle acceleration and scattering \cite{bieber++93},
advection dominated accretion flows, and perhaps solar/stellar wind
acceleration.  Yet key aspects of the physics of
turbulence in magnetized plasmas are poorly understood, in particular
the physics of dissipation at small scales.  The classical scenario of 
magnetohydrodynamic turbulence is thus:  fluctuations in the plasma are driven at some large
'outer' scale and decay by interacting locally in $k-$space.  Eddies at some scale $k$ exchange
energy with eddies at nearby spatial scales, possibly as a three-wave or higher order
interaction \cite{waleffe92, goldreich+sridhar97, ng+batta96},
with the resulting net flow of energy to smaller spatial scales (larger $k$); this cascade of energy occurs
over an 'inertial subrange'  of $k-$space and can be shown to predict a power spectrum that scales as $k^{-5/3}$.  At the smaller scale of the ion thermal gyroradius, $k \rho_i \geq 1$, the ions become demagnetized and the plasma can
no longer behave as a simple fluid; the turbulent energy is then thought to be damped on the thermal plasma
by Landau or transit-time damping.  However, the details of this damping process are not known and
there a few reported measurements in this regime of $k-$space.

Observations of the magnetic spectrum show breakpoints at near $k \rho_i \approx 1$, above
which the spectrum typically becomes harder \cite{beinroth+neubauer81, leamon++98, leamon++99}.  This has been interpreted variously
as evidence of kinetic Alfv\'en waves \cite{leamon++99}, whistler wave dispersion\cite{stawicki++01}, and ion
cyclotron damping of Alfv\'en waves \cite{gary99}.

Here we report the first measured power spectrum of electric fluctuations in solar wind turbulence.  
The inertial subrange is clearly evident and follows the magnetic fluctuation spectrum.  At large
wavenumbers $k \rho_i \geq 1$, the electric spectrum is enhanced.

Data are used from experiments on the Cluster spacecraft.  Cluster flies four
spacecraft, as a controlled tetrahedron, in an inclined orbit with apogee at 19
Earth radii ($R_E$).  From December to May each year, the spacecraft exit
the terrestrial magnetosphere on the dayside and make measurements in the
solar wind.  We use approximately 195 minutes of data during the interval 00:07:00 - 03:21:51 on February 19, 2002, 
when Cluster was at apogee and spent several hours in the ambient solar wind; all of our data is from Cluster spacecraft 4.

The electric field is measured by the Electric Field and Waves experiment (EFW) experiment
 \cite{gustafsson++97}; EFW
is a  double-probe experiment which measures the floating voltage of 8 cm diameter current-biased
spheres extended on 44m wires in quadrature.  These spheres, as well as the spacecraft, are 
illuminated by the Sun and emit photoelectrons which cause the surfaces to charge positive
with respect to the plasma.  The surfaces attract a return current of thermal electrons which provide
the electrical coupling to the plasma.  Systematic variations in this coupling, due to changing illumination
or variations in surface properties and work function, are a large source of background noise in EFW
at the spacecraft spin-period (4 seconds) and harmonics.  This
is discussed more below.  EFW measures the electric field on two orthogonal sensor pairs in the spacecraft
spin-plane at 25 samples/sec.  These two components are rotated into $X$ and $Y$
components in the GSE (geocentric solar ecliptic) coordinate system.  Since the 
GSE $Y$ direction represents the orientation with best symmetry for solar illumination, this
component of the electric field is generally less noisy; we use the GSE $Y$ electric field $E_y$
for all of our analysis.  However, at any given instance, $E_y$ is composed of data from all
four electric field probes, each with slightly different photocoupling to the plasma.  We therefore
apply a finite impulse response (FIR) filter to the data to notch out the primary perturbations
at the spin-tone and some harmonics.

The magnetic field is measured by the FGM instrument \cite{balogh++97}; three-component magnetic
field vectors are sampled at 22 samples/sec (SPS).  In our analysis, we use the GSE $Z$ component
of the magnetic field $B_z$ for reasons that are explained below.  Moments of the solar wind ion distribution (velocity, density, and temperature) are computed from the ion spectrum
measured by the CIS experiment\cite{reme++97}.

Figure 1 shows an overview of the data used in the following analysis; panels a) and b) are wavelet
spectrograms and will be discussed below.  Panel c) shows
the two components of measured electric field $E_x$ and $E_y$ in GSE coordinates.  Panel (d) show the magnetic field data.  Panels
e), f), and g) show the plasma ion density, plasma ion $\beta_i$ (ratio of plasma to magnetic pressure), and
Alfv\'en Mach number.   The average ion beta is $\bar{\beta}_i \approx 5$, average Alfv\'en speed $\bar{v}_A
\approx 40$ km/s, and the average solar wind velocity is $\bar{v}_{sw} \approx$ (-347, 4.9, -32.6) km/s (in GSE
coordinates), over the entire interval.  During the interval between 00:30 and 00:50, the magnetic field is nearly
tangent to the Earth's bow shock (as per a calculation assuming straight field lines \cite{filbert+kellogg79}); however,
Cluster summary plots of electron and plasma wave data show {\em no} evidence of connection to the shock.  All of
our data is ambient solar wind.

To compute power spectra, the electric field data $E_y$ (25 samples/sec) were subsampled onto the 
time tags of the magnetic field data $B_z$ (22 samples/sec) by linear interpolation; a total of exactly $2^{18}$ 
points are used.  The power spectral density (PSD) was computed 
using both Fast Fourier Transform (FFT) and Morelet wavelet \cite{torrence+compo98} schemes.  The FFT was computed as follows:
the data interval was divided into 64 contiguous ensembles of length 4096 (182 seconds);
this gives an inherent bandwidth of $\Delta f \approx 1/186$ Hz.  To minimize spectral leakage, each ensemble was
'whitened' by applying a first-order difference algorithm, the PSD was computed by FFT, then the spectrum
was postdarkened \cite{bieber++93} and divided by the bandwidth of the FFT.  Since
the data is prewhitened, no window function was applied before the FFT.  The electric field spectra
were then 'cleaned' by interpolating over the narrowband spikes resulting from the spin-associated 
signals described above.  A final spectrum was computed as the average of the 64 ensembles.  
Figure 2 shows the FFT power spectra of $E_y$ and $B_z$ (in black).  Wavelet spectra were computed by first producing
the (complex) FFT of $E_y$ and applying the spectral cleaning (interpolation) to the real and imaginary parts, at positive and negative frequencies.
An inverse FFT restores the 'cleaned' signal and a Morelet wavelet spectrogram was computed from this
cleaned $E_y$, as well as the original $B_z$.  The wavelet has 136 log-spaced frequencies; the final wavelet
PSD is computed as the square of the spectrum averaged over time.  The wavelet PSD is also shown in Figure 2
(in red).  The wavelet spectrum extends to lower frequencies than the FFT, which is composed
of ensembles of smaller data intervals; however, these very low frequencies lie below the 'cone
of influence' and are unreliable\cite{torrence+compo98}.  Here we restrict our interpretation to the region
where the FFT and wavelet spectra agree.
The FFT electric spectrum in Figure 2 shows clearly the effect of the notch filters and residual spin-harmonic
spikes.  The wavelet PSD, with its much larger bandwidth, mostly averages over these residual features
although a depression near the notched portion of the spectrum can be seen.  The
FFT and wavelet PSD spectra agree remarkably well for both electric and magnetic fields.

Of course, our (human) scheme of measuring time means little to the solar wind plasma, so 
there is little reason to expect the data to be inherently organized by a power spectrum in Hertz.  Since
the solar wind is super-Alfv\'enic (Figure  1), the phase speed $v_A$ of the 
Alfv\'enic fluctuations is much less than the wind speed itself; hence the measured frequency
spectrum is actually a Doppler-shifted wavenumber spectrum $\omega \approx k v_{sw}$. 
This is often called Taylor's hypothesis and might not be considered to hold at large
wavenumbers, especially if waves are present with phase speeds greater than the solar
wind speed (such as whistler waves).  

As discussed above, it is considered that the fluid-like behavior of the wind
breaks down at near $k \rho_i \approx 1$, therefore $k \rho_i$ is a natural parameter for organization
of the power spectrum.  The top panel of Figure 3 shows the FFT and wavelet power spectra organized by $k \rho_i$, instead of frequency.
For the FFT spectrum, the local values of $|v_{sw}|$, $T_i$, and $|B|$ are used to compute $k = \omega/v_{sw}$
and the thermal ion gyroradius $\rho_i = v_i/\Omega_{ci}$ averaged over each (186 sec) ensemble; the $E_y$ and $B_z$ power
spectra are then interpolated onto a linearly-spaced set of values $k \rho_i \in (0.006, 10)$.  Since solar
wind parameters vary slightly in each ensemble, this also has the effect of smearing (averaging) over
the narrowband interference in the FFT PSD of $E_y$.  The wavelet spectrograms are time-averaged on to 4 second 
intervals and then interpolated onto a set of log-spaced values of $k \rho_i$; panels a) and b) of Figure 1 show
these scaled spectrograms as a function of time.  In Figure 1 a) and b), the fluctuation power has been divided by $k^{-5/3}$
to highlight fluctuations above the average spectrum of the inertial range.  The electric and magnetic wavelet spectrograms
are then averaged to compute the composite spectra in panel a) of Figure 3.

Between $k \rho_i \approx$ 0.015 and 0.45, the wavelet and FFT spectra of electric and magnetic fluctuations show power law behavior with indices of $k^{-1.7}$, which is consistent with the Kolmogorov value of 5/3.  Both $\delta E_y$ and $\delta B_z$ show breakpoints
at near $k \rho_i \approx 0.45$; the magnetic spectrum becomes harder with a index $k^{-2.12}$, while the electric spectrum
becomes softer.  As discussed above, hard magnetic spectra have been observed previously \cite{beinroth+neubauer81,leamon++98}.  Above
$k \rho_i \approx 0.45$, the electric spectrum is power law like $k^{-1.26}$ to $k \rho_i \approx$ 2.5.  Above this second
breakpoint, a exponential $\exp{(-k \rho_i/12.5)}$ better fits the spectrum.  At these higher wavenumbers, the electric field
data is noisy and shows harmonics of the spin tone (as shown above).  To test the validity of this data, we perform two
analyses.  The black dots of panel c)  in Figure 3 show the correlation between the electric and magnetic wavelet power
as a function of $k \rho_i$.  It can be seen that the fluctuations are strongly correlated through the inertial range (with coefficient 
$\approx 1$), remain well-correlated between the two breakpoints $k \rho_i \in$ (0.45, 2) and begin to loose correlation
quickly above the second breakpoint.  A wavelet cross-spectral analysis (between $\delta E_y$ and $\delta B_z$ was also computed; the blue bars show the cross-spectral
coherence, with 1 sigma error bars, also as a function of $k \rho_i$.  Again, $\delta E_y$ and $\delta B_z$ are strongly coherent
through the inertial range and past the first breakpoint.  We conclude that the electric and magnetic spectra
 physical and well-correlated up to the second spectral breakpoint.  Above $k \rho_i \approx$ 2.5 it is difficult to assess the
 quality of the data.  If electrostatic waves are present, there is no expectation of correlation with $\delta B$; however
 in this initial study, we cannot eliminate the possibility of systematic noise at these frequencies.  Additionally, the effects
 of low pass filters on both the EFW and FGM experiments may modify the spectra at these highest ($k \rho_i > 3$) frequencies.

To estimate the phase speed of the fluctuations, we use Faraday's law and compute the ratio of the electric and magnetic spectra.  Since the
electric field measurements are made in the spacecraft (unprimed) frame, we need to Lorentz transform to the plasma
(primed) frame by $\vec{E} = \vec{E}' + \vec{v}_{sw} \times \vec{B}$.  Panel b) of Figure 3 shows the phase speed

\begin{equation}
v_\phi(k \rho_i) = \frac{\delta E'_y (k \rho_i)}{\delta B_z(k \rho_i)} = \frac{\delta E_y (k \rho_i)}{\delta B_z(k \rho_i)} + \overline{v}_x - \overline{v}_z \frac{\overline{B}_x}{\overline{B}_z}
\end{equation}
where $\bar{v}_x, \bar{v}_z, \bar{B}_x$, and $\bar{B}_z$ are the average $x$ and $z$ components of the solar wind
velocity and magnetic field.  The black dots in panel b) are computed from the wavelet spectrum, while the blue line
is computed from the FFT spectrum.  The average Alfv\'en speed $\bar{v}_A \approx$ 40 km/s is shown as a horizontal bar.
Over, and even below, the inertial range $k \rho_i \in$ (0.015, 0.4) the phase speed is consistent with the local Alfv\'en
speed; this is strong evidence of the Alfv\'enic nature of the cascade.  The red curve in panel b) is a fit of the function
$v_0 ~(1 + k^2 \rho_i^2)$ to the FFT curve, where $v_0$ is a free parameter which finds a best fit  at $v_0 \approx 55$ km/s.  This
function approximates the dispersion relation of kinetic Alfv\'en waves.  The cold-plasma whistler
wave phase speed goes as $v_\phi \approx (k \rho_i) ~\beta^{-1/2} ~v_A$ above $\omega > \Omega_{ci}$, i.e. linear with $k \rho_i$,
and would form a much shallower dispersion above $k \rho_i \approx 1$ than is observed 
in panel b) of Figure 3.  This leads us to believe that the Alfv\'en waves in the inertial subrange eventually
disperse as 'kinetic' Alfv\'en waves above $k \rho_i \approx 1$, becoming more electrostatic and eventually damping on the
thermal plasma.  Plasma heating by linear dissipation of kinetic Alfv\'en waves at $\beta \approx 1$ has been studied
in the context of accretion flows\cite{quataert98}.  There it was found that Landau and transit-time damping
contribute to both proton and electron heating at short wavelengths, which is enhanced for higher $\beta$.
  Kellogg \cite{kellogg00} computed the level of
electric field fluctuations required to stochastically thermalize protons to 1 AU in the solar wind; he found that a spectral
density of $E^2 \approx 10^{-11} (V/m)^2 Hz^{-1}$ was sufficient.  This is one order of magnitude less than our observed
levels (Figure 2).  It is, therefore, plausible to conclude that the observed electric spectrum is responsible for isotropizing
the solar wind protons and may be the mechanism by which the solar wind maintains its fluid-like characteristics.

\acknowledgments
Cluster data analysis at UC Berkeley is supported by NASA grants NAG5-11944 and NNG04GG08G.

\newpage

\begin{figure}
\caption{Wavelet and time series data of solar wind turbulence.  From the top-down, 
the five panels show a) the wavelet spectrogram of $E_y$, as a function of $k \rho_i$, 
b) a similar wavelet spectrogram of $B_z$,  c) the $X$ and $Y$ components of the measured electric field,
d) the vector magnetic field, e) plasma ion density, f) plasma ion $\beta$, and g)
the Alfv\'en Mach number.  This entire interval was used for the spectral analysis
of $E_y$ and $B_z$.  The spectral breakpoints called out.}
\label{autonum}
\end{figure}

\begin{figure}
\caption{Power spectral density of electric $\delta E_y$ and magnetic fluctuations $\delta B_z$ as a function of frequency,
computed from FFT (black) and Morlet wavelet (red) algorithms.  The FFT spectrum of electric
field (upper panel) shows the effect of notch filters and residual spin-tone data.  }
\end{figure}

\begin{figure}
\caption{The wavelet (upper) and FFT (lower) power spectra of $E_y$ (green) and $B_z$ (black) 
binned as a function of wavenumber $k \rho_i$ (and offset for clarity) in panel a).  The electric
are multiplied by factor to lie atop the magnetic spectra.  The
spectrum is Kolmogorov $k^{-5/3}$ over the interval $k \rho_i \in (0.015, 0.45)$; a spectral breakpoint
occurs for both $E_y$ and $B_z$ at $k \rho_i \approx 0.45$.  A second breakpoint occurs for electric 
spectrum at $k \rho_i \approx 2.5$ above which the electric spectrum is more exponential.  Panel b) shows the 
ratio of the electric to magnetic spectra in the plasma frame; the average Alfv\'en speed ($\bar{v}_A \approx 40$ km/s)
is shown as a horizontal line.  The red line is a fitted dispersion curve, discussed in the text.  Panel c) shows both
the cross-coherence of $\delta E_y$ with $\delta B_z$ (as blue dots with error bars) and the correlation between
the electric and magnetic power (as black dots). }
\end{figure}
\end{document}